\documentclass[runningheads]{llncs}
\usepackage{graphicx}
\usepackage{mathtools}
\usepackage{amsmath,amsfonts,amssymb}
\usepackage{subcaption}
\usepackage{bm}
\usepackage{multirow}
\usepackage{siunitx}
\usepackage{comment}
\usepackage{listings}
\usepackage{hyperref}
\hypersetup{pdftex=true, colorlinks=true, breaklinks=true, linkcolor=blue, menucolor=blue, citecolor=blue, urlcolor=blue}
\usepackage{xcolor}
\begin{document}
\title{\texttt{PyHexTop}: a compact Python code for topology optimization
using hexagonal elements}

\titlerunning{PyHexTop}

\author{Aditi Agarwal\inst{1}, Anupam Saxena\inst{2} \and
Prabhat Kumar\inst{1}}

\institute{Department of Mechanical and Aerospace Engineering,\\
Indian Institute of Technology Hyderabad,
Telangana 502285, India  \\ 
\and Department of Mechanical Engineering,\\
Indian Institute of Technology Kanpur,
Uttar Pradesh 208016, India
\email{aditi.s.agarwal02@gmail.com},
\email{anupams@iitk.ac.in},
\email{pkumar@mae.iith.ac.in}}
\maketitle              
Published\footnote{This pdf is the personal version of an article whose final publication is available at \href{https://link.springer.com/chapter/10.1007/978-981-96-1158-4_35}{Advances in Multidisciplinary Design, Analysis and Optimization}}\,\,\,in \textit{Advances in Multidisciplinary Design, Analysis and Optimization}, 
\href{https://link.springer.com/chapter/10.1007/978-981-96-1158-4_35}{DOI:10.1007/978-981-96-1158-4\_35} 

\begin{abstract}
Python serves as an open-source and cost-effective alternative to the MATLAB programming language. This paper introduces a concise topology optimization Python code, named ``\texttt{PyHexTop}," primarily intended for educational purposes. Code employs hexagonal elements to parameterize design domains as such elements provide checkerboard-free optimized design naturally. \texttt{PyHexTop} is developed based on the ``\texttt{HoneyTop90}" MATLAB code~\cite{kumar2023honeytop90} and uses the \texttt{NumPy} and \texttt{SciPy} libraries. Code is straightforward and easily comprehensible, proving a helpful tool that can help people new in the topology optimization field to learn and explore. \texttt{PyHexTop} is specifically tailored to address compliance minimization with specified volume constraints. The paper provides a detailed explanation of the code for solving the Messerschmitt-Bolkow-Blohm beam and extensions to solve problems different problems. The code is publicly shared at: \url{https://github.com/PrabhatIn/PyHexTop}.
\end{abstract}
\section{Introduction}
Topology optimization (TO) concerns the best locations to place cavities in a structure design domain by extremising a desired objective with a given set of constraints. This field has undergone extensive research over the past three decades, resulting in numerous research articles that propose various TO algorithms for implementing methods like Solid Isotropic Material with Penalization (SIMP)~\cite{sigmund2013topology}, Level-set method~\cite{van2013level}, evolutionary structural optimization method~\cite{huang2010evolutionary}, and feature-based~\cite{wein2020review,saxena2011topology} approach, etc. Providing educational code in TO is a positive trend initiated by Sigmund~\cite{sigmund200199} with the 99-line MATLAB code. Making such codes publicly available helps the field's growth to different regimes and applications, wherein newcomers find it easy to learn, explore, and extend. Python has emerged as a cost-effective, open-source alternative to the MATLAB programming language and is gaining popularity. Only a few published TO codes using Python interfaces exist~\cite{zuo2015simple,smit2021topology}. This work introduces a standalone Python code called ``\texttt{PyHexTop}" for educational purposes, explicitly targeting newcomers and students. The code uses hexagonal elements to parameterize the design domain. These elements are generated by method outlined in \texttt{HoneyTop90} MATLAB code~\cite{kumar2023honeytop90}. Such elements provide edge connections between two adjacent elements, eliminating checkerboard patterns from the optimized designs~\cite{kumar2023honeytop90,saxena2011topology}.

Python is a dynamic and versatile programming language with great utility in various domains, including TO. Its powerful libraries and frameworks, like \texttt{NumPy} and \texttt{SciPy}, are best suited for scientific computation, data analysis, and numerical optimization. Thus,  Python can be an excellent choice for implementing the TO technique.  The code presented in this paper offers a user-friendly platform that encourages further development and flexibility to implement alternate optimization methods with different underlying physics. 

\begin{figure}[h]
	\begin{subfigure}[t]{0.50\textwidth}
		\centering
		\includegraphics[scale=0.75]{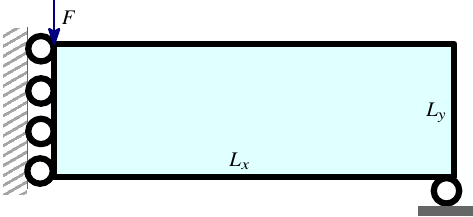}
		\caption{MBB beam}
		\label{fig:MBB}
	\end{subfigure}
	\quad 
	\begin{subfigure}[t]{0.5\textwidth}
		\centering
		\includegraphics[scale=0.75]{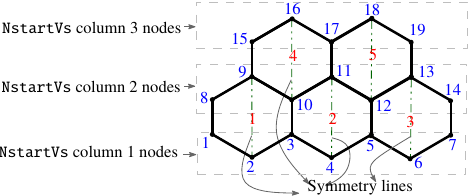}
		\caption{Numbering scheme}
		\label{fig:Schematic2}
	\end{subfigure}
 \begin{subfigure}[t]{0.5\textwidth}
		\centering
		\includegraphics[scale=1]{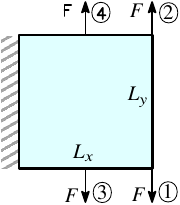}
		\caption{Multi-load cases}
		\label{fig:multiloadDD}
	\end{subfigure}
 \begin{subfigure}[t]{0.5\textwidth}
		\centering
		\includegraphics[scale= 1.25]{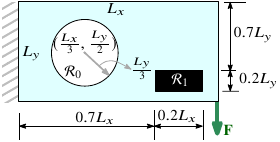}
		\caption{$R_0$: Void region, $R_1$: Solid region}
		\label{fig:PassiveDD}
	\end{subfigure}
	\caption{(\subref{fig:MBB}) A symmetric MBB beam design domain. (\subref{fig:Schematic2}) A schematic for node (blue texts) and element (red texts) numbering scheme~\cite{kumar2023honeytop90}. (\subref{fig:multiloadDD}) Multiple load cases domain. Texts in circles indicate load cases. (\subref{fig:PassiveDD}) Domain with passive regions} \label{fig:Fig1}
\end{figure}
\section{Problem Formulation}
The MBB beam design problem is considered to demonstrate and explain the presented Python code. Fig.~\ref{fig:MBB} depicts the symmetric half-design domain of the beam with the pertinent boundary conditions and external load {F}.

The design domain is parameterized using \texttt{Nelem} hexagonal finite elements~(FEs). Each element is attached with a design variable $\rho_j\in[0,\,1]$. We use the modified SIMP~\cite{sigmund2013topology}, wherein the stiffness matrix of element $j$ is determined as 
\begin{equation}\label{EQ:elementStiff}
\mathbf{k}_j = E(\rho_j)\mathbf{k}_0 = \left(E_\text{min} + \rho_j^p\left(E_0 - E_\text{min}\right)\right)\mathbf{k}_0,
\end{equation}
$E(\rho_j)$, $E_0$ and  $E_\text{min}$ indicate Young's moduli of element $j$,  $\rho_j=1$ (solid), $\rho_j = 0$ (void), respectively.  Material contrast i.e. $\frac{E_\text{min}}{E_0} = 10^{-9}$ is taken~\cite{kumar2023honeytop90}. $\mathbf{k}_0$ is the element stiffness matrix with $E(\rho_j) = 1$. $p$, penalty parameter, is set to 3. We solve the following optimization problem:
\begin{equation} \label{EQ:OPTI} 
\begin{rcases}
\begin{split}
&{\min_{\bm{\rho}}} \quad C({\bm{\rho}}) = \mathbf{u}^\top \mathbf{K}(\bm{\rho})\mathbf{u} = \sum_{j=1}^{\mathtt{Nelem}}\mathbf{u}_j^\top\mathbf{k}_j(\rho_j)\mathbf{u}_j, \quad \bm{0} \leq \bm{\rho} \leq \bm{1},\,\text{with}\\
& (i)\,\,\bm{\lambda}:\,\,\mathbf{K} \mathbf{u} - \mathbf{F} = \mathbf{0},\,(ii)\,\, \Lambda:\,\,\text{g} = \frac{V(\bm{\rho})}{\mathtt{Nelem}\,V^*_f}-1 = \frac{\sum_{j=1}^\mathtt{Nelem}v_j\rho_j}{\mathtt{Nelem}\,V^*_f}-1 \le 0,
\end{split}  
\end{rcases}
\end{equation} 
where $C({\bm{\rho}})$ is compliance of the structure. $\mathbf{u}$ and $\mathbf{F}$ indicate the global displacement and force vectors. $\mathbf{u}_j$ is displacement vector of element $j$. $V$ and $V^*_f*\texttt{Nelem}$ are the optimized designs' total material and desired volumes. $\bm{\rho}$, design vector, contains all $\rho_j$. $\bm{\lambda}$ (vector) and $\Lambda$ (scalar) are the Lagrange multipliers (Eq.~\ref{EQ:OPTI}). $\bm{\lambda = -2\mathbf{u}}$~\cite{kumar2023honeytop90}. Sensitivities of $C$  wrt. $\rho_j$  is $\frac{\partial C}{\partial \rho_j} = -\mathbf{u}_j^\top\frac{\partial \mathbf{k}_j}{\partial \rho_j} \mathbf{u}_j = -p (E_0-E_\text{min})\rho_j^{p-1}\mathbf{u}_j^\top \mathbf{k}_0\mathbf{u}_j$~\cite{kumar2023honeytop90}. Likewise, one finds derivatives of the volume constraint $\frac{\partial \text{g}}{\partial \rho_j} = \frac{v_j}{\mathtt{Nelem}\,V^*_f} = \frac{1}{\mathtt{Nelem}\,V^*_f}$, wherein $v_j =1$ is assumed. The standard optimality criteria~\cite{sigmund200199} is used to solve the optimization problem (Eq.~\ref{EQ:OPTI}). We use sensitivity~\cite{sigmund200199}, density~\cite{bruns2001topology}, and null filterings~\cite{kumar2023honeytop90} to solve the problems presented.

\section{Python Implementation Detail}
This section presents the Python implementation of \texttt{PyHexTop} in detail. We use the method outlined in~\cite{kumar2023honeytop90} to generate the element connectivity and nodal coordinates matrices. The MBB beam is optimized to demonstrate the said code. 

\subsection{Element Connectivity and Nodal Coordinates matrices (lines 15-28)}
Lines 3-8 of \texttt{PyHexTop} are needed to install all the required libraries. 
Let \texttt{HNex} and \texttt{HNey} be the number of hexagonal elements in $x$ and $y$ directions, respectively. Each element consists of six nodes. Each node contains two degrees of freedom (DOFs). The \texttt{Numpy} array \texttt{NstartVs} (line~15) assigns node numbers of elements row by row, starting from the bottom (See~\ref{fig:Schematic2}). \texttt{DOFstartVs} (line~16) stores the DOFs for the symmetrical half quadrilaterals of each hexagonal element, and the order of storage follows the numbering of the nodes, with each node's DOFs arranged sequentially. This array, modified for DOF organization, is horizontally replicated eight times and then vertically replicated $2\times \texttt{HNey}\times \texttt{HNex}$ times. The resulting matrix is named \texttt{NodeDOFs} (line~17). The redundant DOFs related to symmetry are removed using the \texttt{setdiff} \texttt{Numpy} function, and the resulting matrix is stored in \texttt{ActualDOFs} (line~18). The final connectivity matrix \texttt{HoneyDOFs} (line~19) contains organized DOFs relevant to the honeycomb structure, with the specified columns from \texttt{ActualDOFs}. Readers are recommended to refer to~\cite{kumar2023honeytop90} for more detail.

The $y$ coordinates of the nodes for a  honeycomb tessellation are determined on lines 20-22 and stored in array \texttt{Ncyf}. Similarly, the $x$ coordinates are stored on line 24. The matrix \texttt{HoneyNCO} (line 25) records the final nodal coordinates, wherein $x$ and $y$ coordinates are stored in the first and second columns, respectively. 
When \texttt{HNey} is an even number, hanging nodes are found and removed~\cite{kumar2023honeytop90}. \texttt{HoneyDOFs} and \texttt{HoneyNCO} are updated on lines 27 and 28, respectively. 

\subsection{Finite Element Analysis (lines 31-45 and 86-89)}
The total number of elements \texttt{Nelem} and nodes \texttt{Nnode} for the current mesh grid are calculated on line 29. Fixed DOFs are determined and stored in array \texttt{fixeddofs} (line 37). Total DOFs are listed in \texttt{alldofs} (line 38). The array \texttt{freedofs} (line 39) identifies the available DOFs by excluding the fixed DOFs from the set of \texttt{alldofs} by using the \texttt{Numpy} \texttt{setdiff1d} function. 

A sparse force array $\mathbf{F}$ is initialized on line 34 using \texttt{SciPy} function \texttt{csc\_matrix}. The initialization involves assigning a non-zero force value from the \texttt{data} array at the indices specified by \texttt{row} and \texttt{col} vectors. The \texttt{csc\_matrix}  allows for efficient sparse matrix initialization. The displacement vector \textbf{U} is initialized on line 36. On lines 41-44, the rows and column indices of the matrix \texttt{HoneyDOFs} are recorded in vectors \texttt{iK} and \texttt{jK} respectively, using the \texttt{kron} \texttt{Numpy} function. The reshaping and type conversion ensures the indices are appropriately formatted for further use.  \texttt{E0} (line~12) and \texttt{Emin} (line~13) are stored. Elemental stiffness matrix \texttt{KE} (lines 45-56) is evaluated using Wachpress shape functions with plane stress assumptions~\cite{kumar2023honeytop90}. The global stiffness matrix \texttt{K} is determined (line 87). The finite element analysis is performed on lines 86-89 using a sparse linear solver \texttt{spsolve} and \texttt{U} is updated on line 89 with the freed DOFs.

\subsection{Filtering, Optimization and Results Printing (lines 58-81 and and 91-122 )}
Three filtering options are provided: sensitivity filtering $\mathtt{(ft = 1)}$, density filtering $\mathtt{(ft = 1)}$, and null (no) filtering $\mathtt{(ft = 0)}$. The latter is included to illustrate the behavior of hexagonal elements for checkerboard patterns and point connections. We calculate the centroid coordinates of FEs in lines 58-60 (\texttt{Cxx} and \texttt{Cyy}). These vectors are combined in the matrix \texttt{ct} on line 61. 

With \texttt{rfill}, a loop is run to identify neighboring elements within the filter radius, and their indices are recorded in \texttt{I} (line 68). Concurrently, a corresponding index array \texttt{J} is created (line 69). The $x,\, y,$ and distance information of neighboring elements, along with their respective indices, are consolidated into arrays \texttt{newx}, \texttt{newy}, and \texttt{newz}, respectively. Arrays are combined to form the matrix \texttt{DD} (lines 71-74) containing the $x$ and $y$ coordinates and distance of neighboring FEs. The filtering matrix \texttt{HHs} is evaluated on lines 76-77 using \texttt{spdiags} and \texttt{coo\_matrix} \texttt{SciPy} that create a sparse diagonal matrix with indices based on the distances in \texttt{DD}.

 \texttt{volfrac} is used for the initial guess of TO (line~79). $\rho$ is denoted by \texttt{x}. \texttt{xPhys} represents the physical material density vector (line~79). With either $\mathtt{ft = 0}$ or $\mathtt{ft = 1}$, \texttt{xPhys} is equal to \texttt{x}, whereas with $\mathtt{ft = 2}$, it is represented via the filtered density on line 113. \texttt{loop}, \texttt{change}, \texttt{maxiter}, \texttt{dv} and \texttt{move} are defined on line~80. The element compliance \texttt{ce} for each FE is calculated on line 91. \texttt{c} is calculated using material properties and \texttt{xPhys}. The vector \texttt{dc} calculates the sensitivity of \texttt{c} wrt. \texttt{xPhys}, and is later updated as per different \texttt{ft}. 

The optimality criteria for TO is implemented on lines 101-114. \texttt{xOpt} stores the original design variable. \texttt{xUpp} and \texttt{xLow} define the upper and lower bounds of \texttt{x}, respectively. \texttt{OcC} computes the normalized change in \texttt{x}. \texttt{inL} gives an interval for the optimization search.  Vector \texttt{dv} records volume constraint sensitivities. The loop iteratively updates design variables to achieve the desired volume fraction. The convergence and termination conditions are monitored by the vector \texttt{change}. A scatter plot is created to visualize the result using \texttt{xPhys} for color intensity.  
\begin{figure}[h!]
    \centering
       \begin{subfigure}{0.30\textwidth}
        \includegraphics[width=0.99\textwidth, height=0.33\textwidth]{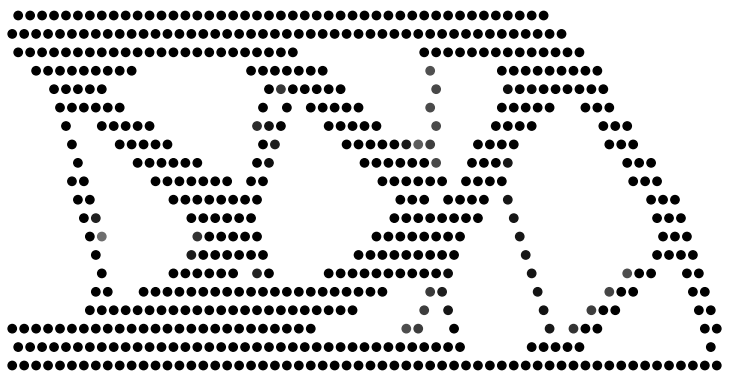}
        \caption{  C = 298.92 }
    \end{subfigure}
    \quad
  \begin{subfigure}{0.30\textwidth}
        \includegraphics[width=0.99\textwidth, height=0.33\textwidth]{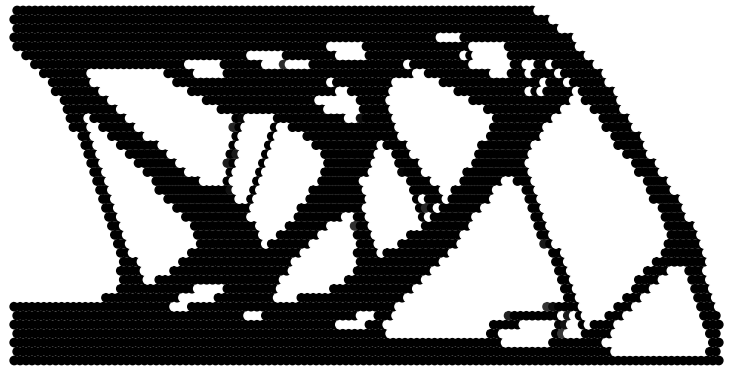}
        \caption{ C = 293.66 }
    \end{subfigure}
    \quad
    \begin{subfigure}{0.30\textwidth}
        \includegraphics[width=0.99\textwidth, height=0.33\textwidth]{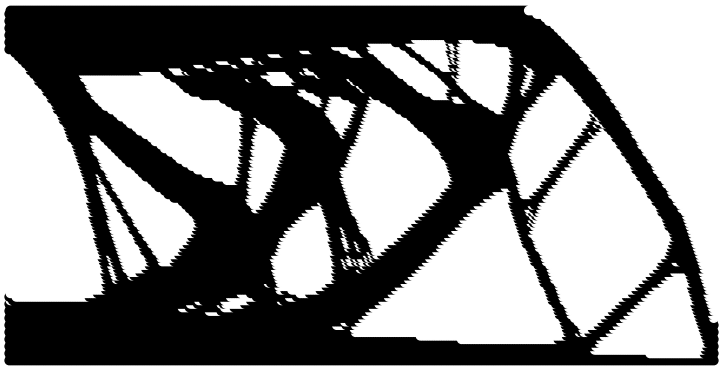}
        \caption{  C = 282.47 }
    \end{subfigure}
    \quad
    \begin{subfigure}{0.30\textwidth}
        \includegraphics[width=0.99\textwidth, height=0.33\textwidth]{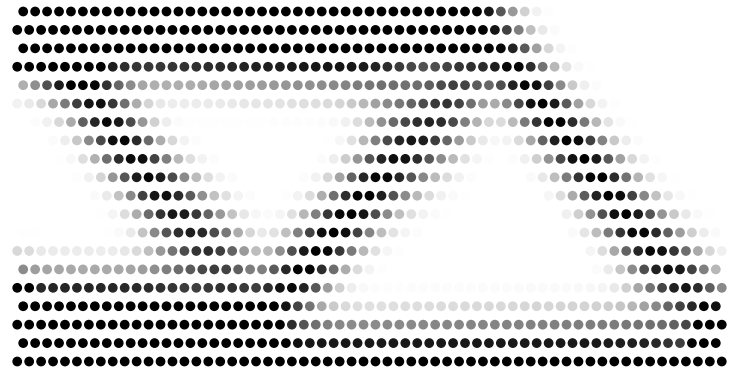}
        \caption{  C = 332.27 }
    \end{subfigure}
    \quad
  \begin{subfigure}{0.30\textwidth}
        \includegraphics[width=0.99\textwidth, height=0.33\textwidth]{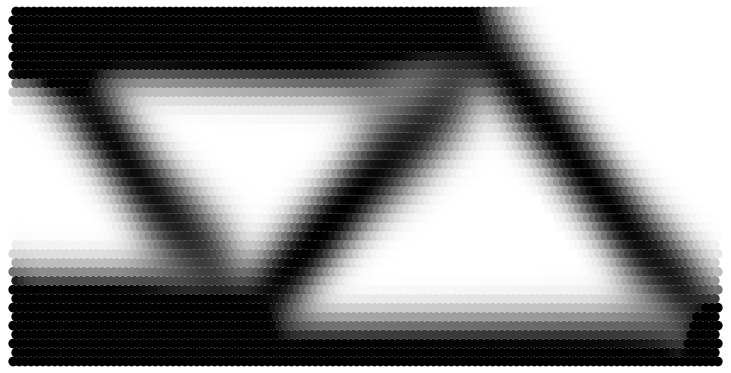}
        \caption{ C = 335.65  }\label{fig:multreslc41}
    \end{subfigure}
    \quad
    \begin{subfigure}{0.30\textwidth}
        \includegraphics[width=0.99\textwidth, height=0.33\textwidth]{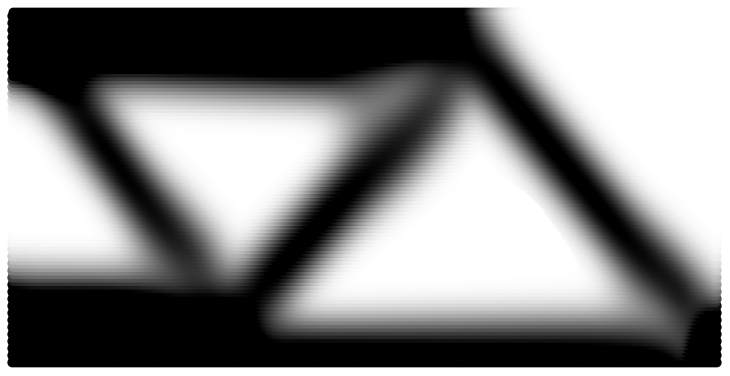}
        \caption{  C = 320.53 }
    \end{subfigure}
    \quad\begin{subfigure}{0.30\textwidth}
        \includegraphics[width=0.99\textwidth, height=0.33\textwidth]{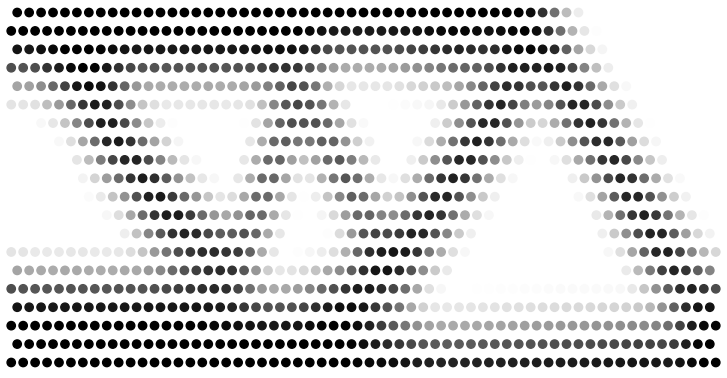}
        \caption{  C = 365.54 }
    \end{subfigure}
    \quad
  \begin{subfigure}{0.30\textwidth}
        \includegraphics[width=0.99\textwidth, height=0.33\textwidth]{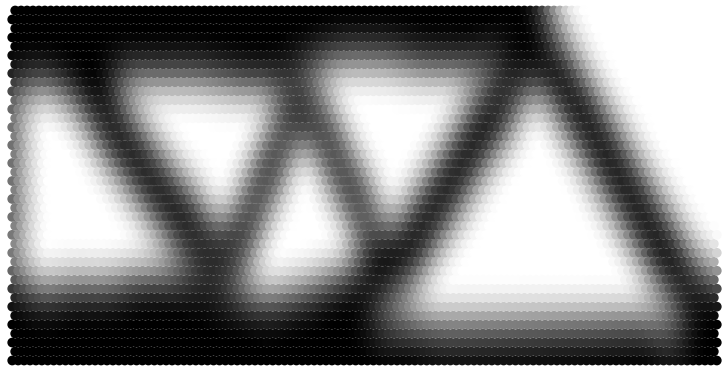}
        \caption{ C = 373.65  }
    \end{subfigure}
    \quad
    \begin{subfigure}{0.30\textwidth}
        \includegraphics[width=0.99\textwidth, height=0.33\textwidth]{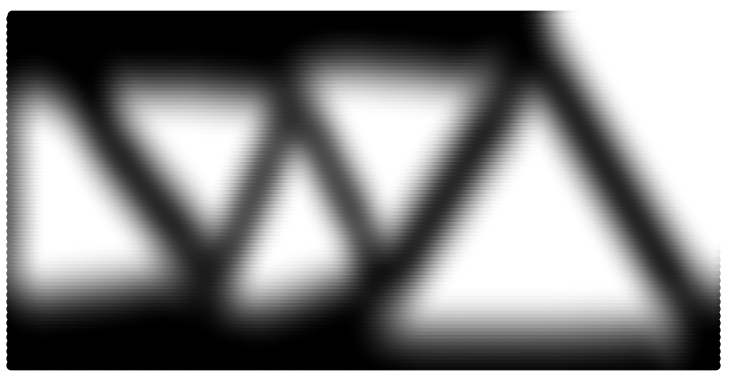}
        \caption{  C = 356.31  }
    \end{subfigure}
    \caption{Optimized MBB beams. The results in the first (a-c), second (d-f), and third (g-i) rows are displayed for ft=0, ft=1, and ft=2 respectively.}\label{Table:T1}
\end{figure}
\subsection{MBB beam optimized results}
\texttt{PyHexTop} code is called as:
\begin{lstlisting}[basicstyle=\ttfamily,breaklines=true,numbers=none]
PyHexTop(HNex, HNey, rfill, volfrac, penal, ft)
\end{lstlisting}
The selected  mesh sizes (HNex $\times$ HNey) are $60 \times 20$, $120 \times 40$ and $300 \times 100$ with the filter radius \texttt{rfill}, $2.4\sqrt{3}$, $4.8\sqrt{3}$, and $12\sqrt{3}$, respectively, i.e., 0.04 times the maximum length. \texttt{volfrac} is set to 0.5, and  \texttt{penal} is set to 3. 

The results in Fig.~\ref{Table:T1} show that designs with $\mathtt{ft=0}$ (1st row) are the best performing and without checkerboard patterns, albeit being mesh-dependent and featuring thin members. To address this, either filtering or length scale constraints~\cite{singh2020topology} can be used. The designs in 2nd and 3rd rows are achieved with $\mathtt{ft=1}$ and $\mathtt{ft=2}$, respectively. These designs eliminate checkerboard patterns and remain mesh-independent, maintaining consistent topology regardless of the mesh sizes.  
\begin{figure}[h!]
    \centering
       \begin{subfigure}{0.2\textwidth}
        \includegraphics[scale = 0.08]{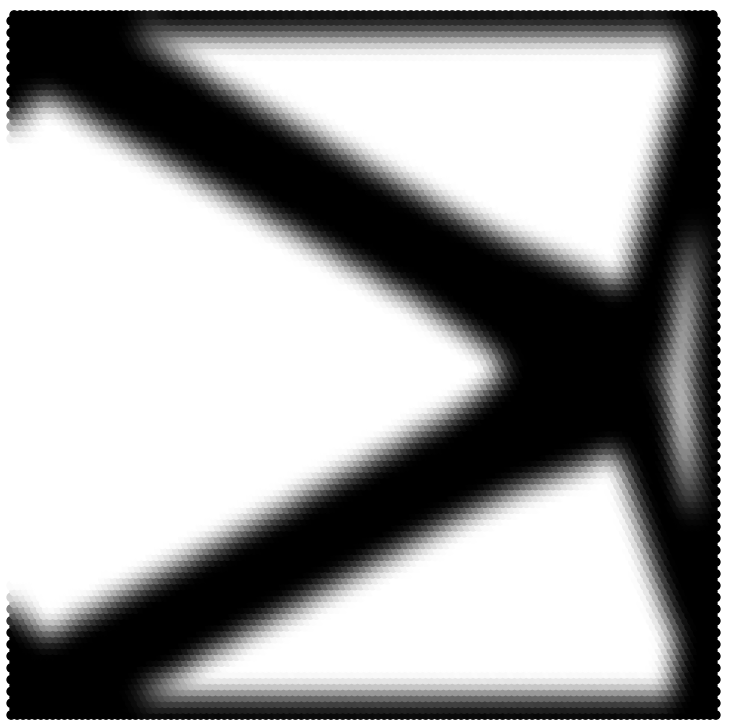}
        \caption{  C = 92.27 }\label{fig:multreslc2}
    \end{subfigure}
 \quad
  \begin{subfigure}{0.2\textwidth}
        \includegraphics[scale = 0.08]{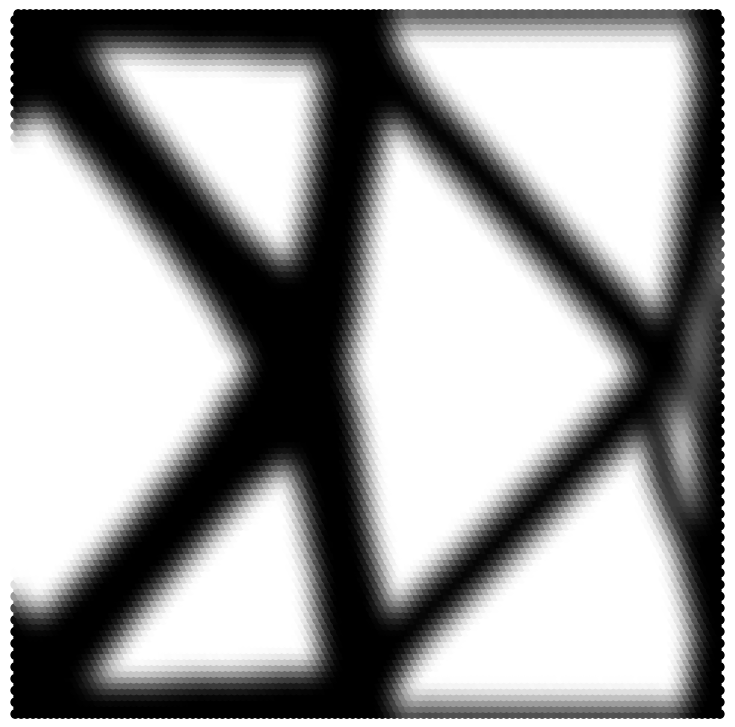}
        \caption{ C = 252.92  }\label{fig:multreslc4}
    \end{subfigure}
    \quad
    \begin{subfigure}{0.2\textwidth}
        \includegraphics[scale =0.15]{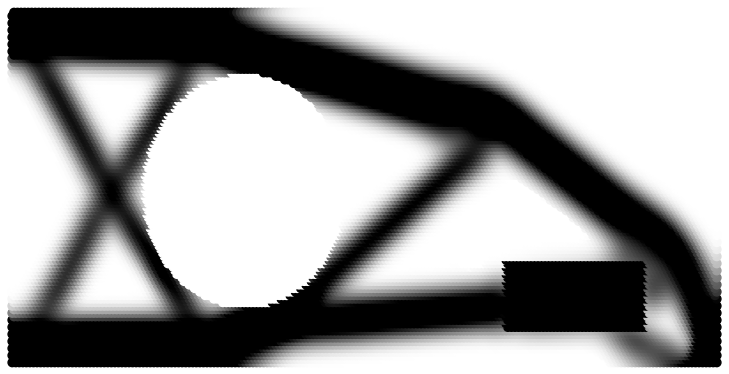}
        \caption{  C = 156.35  }\label{fig:passft1}
    \end{subfigure}
    \quad\quad
    \begin{subfigure}{0.2\textwidth}
        \includegraphics[scale = 0.15]{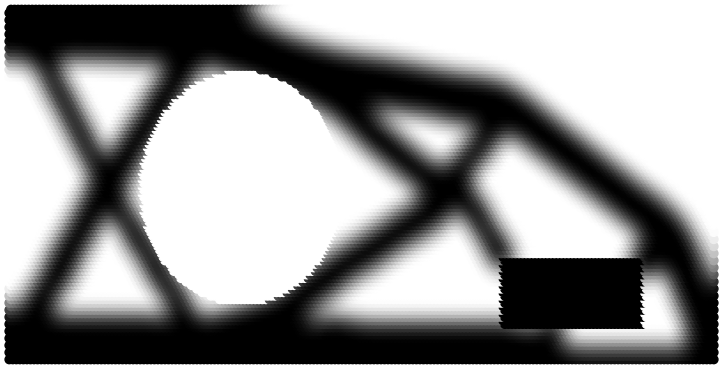}
        \caption{ C = 160.55  }\label{fig:passft2}
    \end{subfigure}
    \caption{Optimized results: multi-load cases   (\subref{fig:multreslc2}) 2-load cases (\subref{fig:multreslc4}) 4-load cases; Passive regions (\subref{fig:passft1}) $\mathtt{ft=1}$ (\subref{fig:passft2}) $\mathtt{ft=2}$}\label{fig:MultiL_Passresults}
\end{figure}
\section{Simple extensions}
This section provides extensions of the Python code for different problems.
\subsection{Multiple loads}
\texttt{PyHexTop} code is modified to solve a cantilever beam design with different multi-load cases  (Fig.~\ref{fig:multiloadDD}). $F$  is adjusted to a column vector per the number of load cases. Accordingly, $U$ is calculated and stored in a two or four-column format.  $C = \sum_{k=1}^{2}\mathbf{U}_k^\top \mathbf{K} \mathbf{U}_k$
where $\boldsymbol{U_k}$ is the displacement for the $k^{th}$ load case. For four load cases, the following modifications are made:
\begin{lstlisting}[basicstyle=\scriptsize\ttfamily,breaklines=true,numbers=none,language=Python]
row = np.array([(2*HNex+1)*2 - 1, 2*Nnode-1, 2*(HNex+1) -1, 2*(HNex + HNey + 2*HNex*HNey) - 1]);col = np.array([0,1,2,3]); data = np.array([-1,1,2,-2]);F = csc_matrix((data, (row, col)), shape=(2*Nnode, 4))

\end{lstlisting}
The \texttt{U} and \texttt{fixeddofs} matrices are adjusted as
\begin{lstlisting}[basicstyle=\scriptsize\ttfamily,breaklines=true,numbers=none,language=Python]
U = np.zeros((2*Nnode, 4));fixeddofs = np.append(2 * (np.arange(1, (2*HNex+1)*HNey+2, 2*HNex+1))-1, 2*(np.arange(1,(2*HNex+1)*HNey+2, 2*HNex+1)))
\end{lstlisting}
The following changes are done to determine objective  and sensitivity:
\begin{lstlisting}[basicstyle=\scriptsize\ttfamily,breaklines=false,numbers=none,language=Python]
c=0;dc = 0
for i in range(0,(F.shape[1])):
    Ui = U[:,i];ce = np.sum((Ui[HoneyDOFs-1].dot(KE))*(Ui[HoneyDOFs-1]),1)
    c = c+sum((Emin+xPhys**penal*(E0-Emin))* (ce.reshape((-1, 1)))).sum()
    dc = dc - penal*(E0 - Emin) * xPhys ** (penal-1)*(ce.reshape((-1, 1)))
\end{lstlisting}
The code is called as below.
\begin{lstlisting}[basicstyle=\scriptsize\ttfamily,numbers=none]
PyHexTop(120, 120,4 * np.sqrt(3), 0.4, 3, 1)
\end{lstlisting}
 The optimized designs for 2-load and 3-load cases are depicted in Figs.~\ref{fig:multreslc2} and~\ref{fig:multreslc4}. 
\subsection{Passive design domains}
A problem~\cite{kumar2023honeytop90} with passive (non-design) regions is shown in Fig.~\ref{fig:PassiveDD}. These modifications are made to code for the problem. The force and fixed DOFs vectors are changed to
\begin{lstlisting}[basicstyle=\scriptsize\ttfamily,numbers=none]
row = np.array([2*(2*HNex+1)-1]);col = np.array([0]);data = np.array([-1])   
F = csc_matrix((data, (row, col)), shape=(2*Nnode, 1))
\end{lstlisting}
\begin{lstlisting}[basicstyle=\scriptsize\ttfamily,numbers=none]
fixeddofs = np.append(2 * (np.arange(1, (2*HNex+1)*HNey+2,  2*HNex+1))-1, 2*(np.arange(1, (2*HNex+1)*HNey+2, 2*HNex+1)))
\end{lstlisting}
\noindent Array \texttt{RSoVo}, is initialised to \texttt{zeros(Nelem,1)} after line 65 to store information related to the regions $\mathbf{R_1}$ (solid) and $\mathbf{R_0}$ (void). \texttt{RSoVo} is then modified as follows (line 74-77)

\begin{lstlisting}[basicstyle=\scriptsize\ttfamily,numbers=none]
if (np.sqrt((ct[j,0]-np.max(ct[:,0])/3)**2+(ct[j,1]-np.max(ct[:,1])/2)**2)<np.max(ct[:,1])/3): 
RSoVo[j] = -1   #non-design void region
elif (ct[j,0]>0.7*np.max(ct[:,0]) and ct[j,0]<0.9* np.max(ct[:,0]) and ct[j,1]>0.1*np.max(ct[:,1]) and ct[j,1]<0.3* np.max(ct[:,1])):
RSoVo[j] = 1    #non-design solid region
\end{lstlisting}
FEs with $\texttt{RSoVo} = -1$ or $\texttt{RSoVo} = 1$ are passive elements. Vector \texttt{act} is included after line 77 to store active FEs. \texttt{NVe} and \texttt{NSe} are arrays for the $\mathbf{R_0}$ and $\mathbf{R_1}$ regions respectively.

\begin{lstlisting}[basicstyle=\scriptsize\ttfamily,numbers=none]
NVe,NSe = np.where(RSoVo == -1), np.where(RSoVo == 1)
act=np.setdiff1d(np.arange(1,Nelem + 1).reshape(-1, 1),np.union1d(NVe,NSe))
x [act-1] = (volfrac * (Nelem - len(NSe)) - len(NVe))/len(act)
\end{lstlisting}
\texttt{x[act-1]} contains initialized design variables per \texttt{act}.
The sensitivities of the objective and constraints wrt. the passive variables are zero. Only active sensitivities are taken on line 113 for performing optimization. The following code is added after density filtering on line 103 for accounting non-design regions.

\begin{lstlisting}[basicstyle=\scriptsize\ttfamily,numbers=none]
xPhys[RSoVo == 1] = 1; xPhys[RSoVo == -1]= 0
\end{lstlisting}
The code is called as below and optimized designs for \texttt{ft=1} and \texttt{ft=2} are displayed in Figs.~\ref{fig:passft1} and~\ref{fig:passft2}, respectively. 
\begin{lstlisting}[basicstyle=\scriptsize\ttfamily,numbers=none]
PyHexTop(200,100, 5.6* np.sqrt(3), 0.4, 3, ft)
\end{lstlisting}

\section{Closure}
This paper provides a Python code, named \texttt{PyHexTop}, for TO with hexagonal elements. Such elements provide edge connectivity between two juxtaposed elements, thus helping achieve checkerboard-free optimized designs. The code aims to facilitate newcomers to learn, explore, and develop codes for various applications. Efficient use of Python's \texttt{NumPy} and \texttt{SciPy} libraries allows to generate nodal coordinates and element connectivity matrix per~\cite{kumar2023honeytop90}. Stiffness matrix of elements is calculated using Wachpress shape functions. \texttt{PyHexTop} employs the optimality criteria approach to solve compliance minimization problems.

\texttt{PyHexTop} and its extensions are discussed, offering a comprehensive understating of the code's implementation. Moreover, simplicity and flexibility of \texttt{PyHexTop} open up several possibilities for future research and developments, e.g., problems with stress/buckling constraints and design-dependent loads~\cite{kumar2023TOPress,kumar2023sorotop}.

\end{document}